\documentclass[%
 aip,
 jmp,%
 amsmath,amssymb,
reprint,%
author-numerical,%
showkeys,
]{revtex4-1}

\usepackage{graphicx}
\usepackage{dcolumn}
\usepackage{bm}

\def\eps{\varepsilon}

\begin{document}


\title[On finite-size Lyapunov exponents in multiscale systems]{On finite-size Lyapunov exponents in multiscale systems}

\author{Lewis Mitchell}
\email{Lewis.Mitchell@uvm.edu}
\affiliation{School of Mathematics and Statistics, University of Sydney, NSW 2006, Australia.}
\affiliation{Department of Mathematics and Statistics, University of Vermont, Burlington, VT 05401, USA.}
\author{Georg A. Gottwald}
\email{georg.gottwald@sydney.edu.au}
\affiliation{School of Mathematics and Statistics, University of Sydney, NSW 2006, Australia.}

\date{\today}

\begin{abstract}
We study the effect of regime switches on finite size Lyapunov exponents (FSLEs) in determining the error growth rates and predictability of multiscale systems. We consider a dynamical system involving slow and fast regimes and switches between them. The surprising result is that due to the presence of regimes the error growth rate can be a non-monotonic function of initial error amplitude. In particular, troughs in the large scales of FSLE spectra is shown to be a signature of slow regimes, whereas fast regimes are shown to cause large peaks in the spectra where error growth rates far exceed those estimated from the maximal Lyapunov exponent. We present analytical results explaining these signatures and corroborate them with numerical simulations. We show further that these peaks disappear in stochastic parametrizations of the fast chaotic processes, and the associated FSLE spectra reveal that large scale predictability properties of the full deterministic model are well approximated whereas small scale features are not properly resolved. 
\end{abstract}


\date{\today}

\maketitle


\begin{quotation}
The atmosphere and the climate system are inherently complex multiscale systems with processes spanning spatial scales from millimetres to thousands of kilometres, and temporal scales from seconds to millennia. It is a formidable challenge to find consistent and effective reduced dynamical equations for the ``slow" and ``large" degrees of freedom with predictive power. An important question is how to measure the radius of predictability in such a multiscale system. One such measure is the maximal Lyapunov exponent $\lambda_{\rm max}$. A generic situation is that the fast degrees are strongly chaotic, causing the Lyapunov exponent of the whole system to be large, indicating poor predictability. However, the large scale slow behaviour can still be forecast with reasonable accuracy for times much longer than ${\cal{O}}(1/\lambda_{\rm max})$; for example, weather can be forecast on time scales above those expected from small-scale instabilities such as convection \cite{Lorenz69}. This is usually due to the small-scale instabilities growing faster but becoming nonlinearly saturated at a much smaller level than large scale instabilities. 
\end{quotation}


\section{Introduction}

In a series of papers \citet{Aurell97} and \citet{Boffetta98,Boffetta98b} introduced the {\em Finite Size Lyapunov Exponent} (FSLE) to extend the idea of measuring the divergence of nearby trajectories to resolve predictability measures for processes developing on various scales. FSLEs have been successfully used in recent years to study mixing and transport problems in lakes \cite{Karolyi10} and ocean currents \cite{HernandezCarrasco11}, and to study meso-scale and sub-mesoscale filamentary processes in the surface circulation \cite{DOvidio04,DOvidio09,TewKai09}.

The study of error growth rates has also led independently \cite{TothKalnay93,TothKalnay97} to the development of {\em breeding vectors} to produce optimal perturbations for ensemble forecasting. The average growth rate of these vectors is closely related to the FSLE \cite{Boffetta02,Pena04}. Besides studying predictability \cite{CaiKalnayToth03,DerembleEtAl09} optimal finite-size perturbations have been used in ensemble forecasting \cite{TothKalnay93,TothKalnay97} and data assimilation \cite{Kalnay} where they should represent the directions of growing analysis errors, which are again scale dependent. Hence, the question of how error growth rates depend on initial error amplitude is integral to the generation of perturbation ensembles for ensemble forecasting and data assimilation.

The particular aspect we address here is how to interpret the FSLE spectrum in a multiscale dynamical system which involves abrupt switches between regimes. We study a system which involves both slow and fast regimes. The meaning of ``slow" depends on the context; synoptic weather systems such as high and low pressure fields are slow when compared to gravity waves, the buoyancy oscillations of stratification surfaces of the atmosphere. However, weather itself is fast when it comes to climate modelling in coupled ocean-atmosphere models, where the ocean evolves on a much slower time scale than the atmosphere.  

Slow weather regimes have long been associated with climate. In the atmosphere they can be associated with zonal and blocked flows dominating weather on time scales up to several weeks \cite{CharneyDeVore79,LegrasGhil85}. Atmospheric slow regimes are responsible for low-frequency variability of planetary scale dynamics \cite{BranstatorBerner05,Majda06}, the Arctic Oscillation and North-Atlantic Oscillation (NAO), the dominant pattern of atmospheric variability over the Atlantic \cite{Kondrashov04}. In the ocean slow regimes have been associated with low-frequency variability of the thermohaline circulation and ENSO \cite{Dijkstra}. On paleoclimatic scales slow regimes distinguish glacial and interglacial periods \cite{Ditlevsen99, KwasniokLohmann09}.

Fast regimes have recently been considered to be highly relevant for atmospheric and climatic variability, and may determine predictability of dominant slower processes. For example, synoptic weather events such as Rossby-wave breaking and high-latitude blocking episodes with life times of 5-10 days are important for the NAO and may give rise to its low-frequency variability on interannual and longer time scales \cite{Woolings08,Greatbatch00}. Similarly, the ENSO variability on time scales of seasons to years can be produced or maintained by faster subseasonal long-lived transient westerly wind bursts and the Madden-Julian Oscillation with life times of 30-90 days 
\cite{ZavalaGaray03,Frauen10,RoulstonNeelin00}. On the mesoscale, fast mesovortices with life times of a few hours can dampen the intensification of hurricanes by mixing heat and momentum \cite{Nguyen11}. 

We investigate a low-dimensional toy model describing one slow metastable degree of freedom coupled to a fast chaotic system. To study the influence of fast regimes, the slow variable will be coupled to two types of fast dynamics, the Lorenz-63 system which involves regimes and the R\"ossler system which does not. The deterministic system under consideration is amenable to stochastic singular perturbation theory (for both types of fast dynamics) which allows us to effectively describe the slow dynamics in a dimension-reduced stochastic model, which supports the same slow regimes. 
We will show that the presence of slow metastable states causes the FSLE spectrum to have a pronounced trough at large scales. Fast regimes, on the other hand, may cause the FSLE spectrum to exhibit large peaks. We develop a quantitative theory which explains both phenomena.

The paper is organized as follows. In Section~\ref{sec:FSLE} we briefly introduce the FSLE. The toy model under consideration is introduced in Section~\ref{sec:model}. Numerically obtained FSLE spectra of the model are presented in Section~\ref{sec:num}. The signatures of slow metastable states on the FSLE spectra is explained analytically in Section~\ref{sec:slowregime} using the multimodal probability density function of the slow variables. The signature of fast regimes on the FSLE spectra is quantitatively explained by means of a heuristic argument in Section~\ref{sec:fastregime}. We conclude with a discussion in Section~\ref{sec:discussion}.


\section{Finite Size Lyapunov Exponents}
\label{sec:FSLE}
The finite size Lyapunov exponent (FSLE) introduced by \citet{Aurell97} and \citet{Boffetta98,Boffetta98b} measures the growth rate of a perturbation of finite size $\delta$. The FSLE $\lambda(\delta)$ is defined as
\begin{align}
\lambda(\delta) = \left< \frac{1}{T_r(\delta)} \right>_\mu\ln r = \frac{1}{\left< T_r(\delta) \right>_{\rm ens}} \ln r \; ,
\label{e.FSLE}
\end{align}
where $T_r(\delta)$ is the time taken for a perturbation of size $\delta$ to grow by an amplification factor $r$, which we take to be $r = 1.1$ throughout. The first average $< \cdot >_\mu$ is taken over the invariant measure of the dynamics which is approximated by the ensemble average $< \cdot >_{\rm ens}$ over many realizations. To compute the FSLE spectrum, i.e. $\lambda$ as a function of $\delta$, numerically, two trajectories are created starting with an initial separation $\delta_0$, and the separation $\delta$ is measured as the trajectories diverge over time. The perturbations $\delta$ are assumed to be already aligned with the most unstable direction, which is guaranteed by initializing each realization with a sufficiently small initial perturbation size $\delta_0$. Note that the small-scale FSLE with $\delta\to 0$ corresponds to the maximal Lyapunov exponent $\lambda_{\rm max}$.


\section{The model}
\label{sec:model}
We study multiscale systems of the form
\begin{align}
\frac{dx}{dt} &= ax(b^2-x^2) + \frac{1}{\varepsilon}f(y) 
\label{eqn:x0}\\
\frac{dy}{dt}&= \frac{1}{\varepsilon^2}g(y)\; , 
\label{eqn:y}
\end{align}
in which a slow degree of freedom $x\in\mathbb{R}$ describes an overdamped degree of freedom in a double-well potential 
\begin{align}
V(x) = a\frac{x^4}{4} - ab^2\frac{x^2}{2}\; ,
\label{V}
\end{align}
which is driven by a fast chaotic process $y\in \mathbb{R}^3$. The parameter $b$ controls the location of the slow metastable states near $x^* = \pm b$ and their separation $2b$. The height of the potential barrier $\Delta V(x)=ab^4/4$ is controlled by both $a$ and $b$. Unless otherwise specified, we set $a=b=1$, and $\varepsilon^2=0.01$.\\

We consider three cases: where the fast subsystem is given by A.) the chaotic Lorenz-63 system, B.) the chaotic R\"ossler system and C.) a reduced stochastic system which we derive to describe the statistics of the effective slow dynamics only. The slow $x$-dynamics supports slow metastable states near $x^* = \pm b$ in all three cases. However, only the Lorenz-63 system supports fast regimes.

\begin{figure}[]
\centering
\includegraphics[width = \columnwidth]{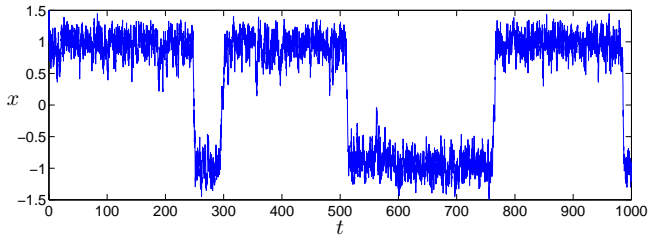}
\caption{Sample trajectory of the metastable slow variable $x$ calculated from the system (\ref{eqn:x})-(\ref{eqn:y3}).}
\label{fig:sampletraj}
\end{figure}

Figure~\ref{fig:sampletraj} shows a sample trajectory of the slow variable $x$ for the Lorenz-driven system (see (\ref{eqn:x})--(\ref{eqn:y3}) below) which clearly shows how the fast chaotic process causes the slow variable to switch between regimes centred around $x^\star=\pm1$. Simulations of the R\"ossler-driven system and of the reduced stochastic system exhibit qualitatively similar behaviour.

%


\subsection{Fast Lorenz-63 subsystem with regimes}
\label{sec:L63}
We consider the multiscale model (\ref{eqn:x0})-(\ref{eqn:y}) when the slow dynamics is driven by a fast Lorenz-63 subsystem
\begin{align}
	\frac{dx}{dt} &= x-x^3 +\frac{k}{\varepsilon}y_2 \label{eqn:x}\\
	\frac{dy_1}{dt}&= \frac{10}{\varepsilon^2}(y_2-y_1) \label{eqn:y1}\\
	\frac{dy_2}{dt}&= \frac{1}{\varepsilon^2}(28y_1-y_2-y_1y_3)\\
	\frac{dy_3}{dt}&= \frac{1}{\varepsilon^2}(y_1y_2-\frac{8}{3}y_3)\label{eqn:y3}\; .
\end{align}
This system was introduced by \citet{Givonetal04} and has been further analyzed by \citet{Mitchell11b}. We use here $k=4/90$ which produces an autocorrelation decay time of the slow variable $\tau_{\rm corr} \approx 208$ time units. The maximal Lyapunov exponent is estimated as $\lambda_{\rm max} \approx 97.4$ and scales with $\varepsilon^2$, the time scale of the fast dynamics. This exemplifies the discrepancy between large scale predictability as measured by $\tau_{\rm corr}$ and the inverse of the maximal Lyapunov exponent.\\

The fast dynamics contains regimes consisting of the two respective lobes of the butterfly attractor and abrupt switches between them. We remark that strictly the metastable states of the Lorenz-63 system do not consist of the lobes of the butterfly attractor, but involve parts of the attractor from both lobes, separated by the stable manifold of the lowest period symmetric unstable periodic orbit \cite{FroylandPadberg09}. Here however, we use the common terminology of {\em regimes}, meaning the lobes of the butterfly attractor.\\


\subsection{Fast R\"ossler subsystem with no regimes}
\label{sec:Roessler}
Further, we consider the multiscale model (\ref{eqn:x0})-(\ref{eqn:y}) when the slow dynamics is driven by a fast R\"ossler subsystem
\begin{align}
	\frac{dx}{dt} &= x-x^3 +\frac{k}{\varepsilon}(y_2-\bar{y}_2) \label{eqn:x_r}\\
	\frac{dy_1}{dt}&= \frac{1}{\varepsilon^2}(-y_2-y_3) \label{eqn:y1_r}\\
	\frac{dy_2}{dt}&= \frac{1}{\varepsilon^2}(y_1 + 0.432 y_2)\label{eqn:y2_r}\\
	\frac{dy_3}{dt}&= \frac{1}{\varepsilon^2}(2 + y_3(y_1 - 4))\label{eqn:y3_r}\; .
\end{align}
Unlike the Lorenz-63 system, the fast R\"ossler system has only one unstable fixed point and does not support regimes. The coupling is chosen so that the forcing has mean zero; the mean of the driving R\"ossler variable was estimated as $\bar{y}_2 \approx -0.939$ from a long trajectory. We chose the coupling parameter $k = 0.525$, which corresponds to an autocorrelation decay time of the slow variable $\tau_{\rm corr} \approx 234$ time units, comparable to that calculated for the Lorenz-driven system. The maximal Lyapunov exponent for the fast subsystem is measured to be $\lambda_{\rm max} = 10.12$, scaling again with $\varepsilon^2$. \\ 


\subsection{Reduced homogenized stochastic slow dynamics}
\label{sec:homo}

The multiscale system (\ref{eqn:x0})-(\ref{eqn:y}) can be reduced using stochastic singular perturbation theory (homogenization) \cite{PavliotisStuart,MelbourneStuart11}
in the case that the fast dynamics is mixing and the average of the slow vectorfield $f(y)$ over the ergodic measure induced by the fast process vanishes. Ergodicity and the mixing property have been rigorously proven for the chaotic Lorenz-63 system \cite{Tucker99,Luzzatto05}, and numerical simulations suggest that they exist for the R\"ossler system as well. The centering condition of the vanishing average of the fast vectorfield $f(y)$ is automatically satisfied for the Lorenz-driven system (\ref{eqn:x})-(\ref{eqn:y3}) since the average of $y_2$ is zero, and by construction for the R\"ossler-driven system (\ref{eqn:x_r})-(\ref{eqn:y3_r}) for sufficiently accurate numerical estimates of the average ${\bar{y}}_2$. 

In stochastic homogenization the fast chaotic degrees of freedom are parametrized by a stochastic process, provided the fast processes decorrelate rapidly enough that the slow variables experience the sum of uncorrelated fast dynamics during one slow time unit. According to the (weak) Central Limit Theorem this corresponds to approximate Gaussian noise. Applying homogenization the following reduced stochastic model can be deduced for the slow $x$-dynamics
\begin{equation}
\frac{dX}{dt} = X (1- X^2) + \sigma \frac{dW}{dt} \label{eqn:climate}
\end{equation}
with one-dimensional Wiener process $dW$, and where $\sigma$ is given by the integral of the autocorrelation function of the fast $y_2$ variable with
\begin{align}
\frac{\sigma^2}{2} = 
k^2
\int_0^\infty 
\{ \lim_{T \rightarrow \infty} \frac{1}{T} \int_0^T y_2(s)y_2(t+s)ds \}\,dt\; .
\label{e.sigma2}
\end{align}
In the case of the Lorenz-driven system (\ref{eqn:x})-(\ref{eqn:y3}) the diffusion coefficient is estimated as $\sigma^2 = 0.113$ from a long-time trajectory, for which the decay time of the autocorrelation function is $\tau_{\rm corr} \approx 222$ time units. For details the reader is referred to  Refs. \onlinecite{Givonetal04,Mitchell11b}.\\

\begin{figure}[]
\centering
\includegraphics[width = \columnwidth]{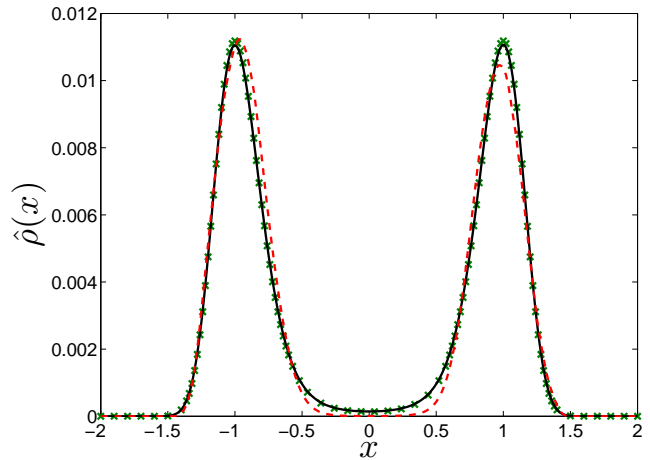}
\caption{Empirical density function of the slow variable $x$ calculated from a long time simulation with $T=10^6$ of the system (\ref{eqn:x0})-(\ref{eqn:y}). Empirical density functions are shown for the Lorenz-driven system (\ref{eqn:x})-(\ref{eqn:y3}) (continuous black line), for the R\"ossler-driven system (\ref{eqn:x_r})-(\ref{eqn:y3_r}) (dashed line, online red) and for the reduced stochastic system (\ref{eqn:climate}) with $\sigma^2=0.113$ (crosses, online green).}
\label{fig:pdf}
\end{figure}

Whereas the invariant ergodic probability density functions can only be numerically estimated for the deterministic equations (\ref{eqn:x})-(\ref{eqn:y3}) and (\ref{eqn:x_r})-(\ref{eqn:y3_r}), it is readily analytically determined for the stochastic gradient Langevin equation (\ref{eqn:climate}) as the unique stationary solution $\hat{\rho}(x)$ of its associated Fokker-Planck equation
\begin{align*}
\frac{\partial}{\partial t} \rho(x) = \frac{\partial}{\partial x}\left(\frac{dV}{dx}\,\rho\right) + \frac{\sigma^2}{2}\frac{\partial^2}{\partial {x^2}}\rho\; .
\end{align*}
We find 
\begin{align}
\hat{\rho}(x)=\frac{1}{Z}e^{-\frac{2}{\sigma^2}V(x)}
\quad {\rm{with}} \quad 
Z=\int_{-\infty}^\infty e^{-\frac{2}{\sigma^2}V(x)}dx\; ,
\label{e.rhohat}
\end{align}
which is depicted in Figure~\ref{fig:pdf}.

In Figure~\ref{fig:pdf} we show the empirical probability density functions of the slow variable $x$ from the Lorenz-driven and R\"ossler-driven systems (\ref{eqn:x})-(\ref{eqn:y3}) and (\ref{eqn:x_r})-(\ref{eqn:y3_r}) obtained from long-time numerical simulations, as well as the exact density function (\ref{e.rhohat}) for the stochastic system (\ref{eqn:climate}) with $\sigma^2 = 0.113$. For this value of $\sigma^2$ the stochastic reduced system approximates the statistics of the full dynamics of the Lorenz-driven system (\ref{eqn:x})-(\ref{eqn:y3}) very well \cite{Mitchell11b}. This is reflected in the close correspondence of the respective empirical density functions. The probability density functions clearly show a bimodal structure indicative of the (slow) metastable nature of the $x$-dynamics as already encountered in Figure~\ref{fig:sampletraj}. The R\"ossler-driven system, however, exhibits a slight asymmetry of the empirical probability density function with one maximum larger than the other. This is caused by the only approximate numerical estimate of ${\bar y}_2$. When numerically simulating the equation for the slow degree of freedom (\ref{eqn:x_r}) for the slow degree of freedom of the R\"ossler driven system we approximate the average of $y_2$ and may write 
\[
\frac{dx}{dt} = x-x^3 + \frac{k}{\varepsilon}(\bar{y}_2-\hat{y}_2) +  \frac{k}{\varepsilon}(y_2-\bar{y}_2)\; ,
\]
where ${\hat y}_2$ is the numerically estimated average of $y_2$ and ${\bar{y}}_2$ the true average of $y_2$. Provided $\alpha=\bar{y}_2-\hat{y}_2 = {\mathcal{O}}(\varepsilon)$ the centering condition (i.e. the vanishing of the average of the $1/\varepsilon$-part of the slow vectorfield over the ergodic measure induced by the fast dynamics) is satisfied. Hence the corresponding reduced stochastic equation (\ref{eqn:climate}) is modified by an additional small term $\alpha/\varepsilon$. This term modifies the potential to $V(x) = a x^4/4 - ab^2 x^2/2 + \alpha x$ causing the slight asymmetry in the probability density function as seen in Figure~\ref{fig:pdf}.

%
%


\section{Non-monotonicity of FSLE spectra for systems involving regimes: numerical simulations}
\label{sec:num}

We now numerically determine the FSLE spectra $\lambda(\delta)$ using only data of the slow $x$-variable for our three cases; (\ref{eqn:x})-(\ref{eqn:y3}) with slow and fast regimes, (\ref{eqn:x_r})-(\ref{eqn:y3_r}) with only slow regimes and (\ref{eqn:climate}) with only slow regimes. In all simulations we initialize the estimation with an initial perturbation size of $\delta_0=10^{-9}$ and average the FSLE spectra over 5000 realizations.\\

\begin{figure}[]
\centering
\includegraphics[width = \columnwidth]{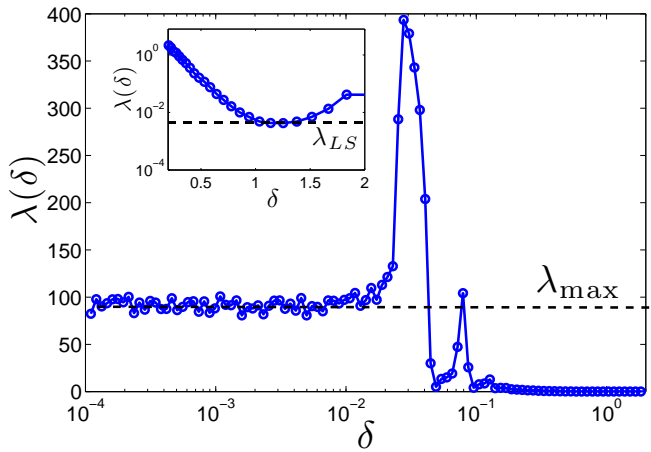}\\
\caption{FSLE $\lambda$ as a function of perturbation size $\delta$ for the Lorenz-driven system (\ref{eqn:x}-\ref{eqn:y3}) with slow and fast regimes. The inset shows a zoom for large scale disturbances.}
\label{fig:FSLE}
\end{figure}

Figure~\ref{fig:FSLE} shows the FSLE as a function of perturbation size $\delta$ for the Lorenz-driven model (\ref{eqn:x})-(\ref{eqn:y3}). For small values of the perturbation size $\delta$ we find the well known plateau corresponding to the maximal Lyapunov exponent which was estimated to be $\lambda_{\rm max} \approx 97.4$ (for $\varepsilon^2 = 0.01$) \cite{Aurell97,Boffetta98,Boffetta98b}.\\ Surprisingly, the FSLE spectrum contains several peaks, notably near $\delta = 0.0278$ and $\delta = 0.0790$, in stark contrast to the behaviour reported in Refs. \onlinecite{Aurell97,Boffetta98,Boffetta98b}. Note that the FSLE at the first peak is much larger than the maximal Lyapunov exponent $\lambda_{\rm max}$ suggesting a far greater loss of predictability at those scales. Initial error amplitudes of those sizes experience much stronger growth than the eigendirections corresponding to the maximal Lyapunov exponent.\\ Interestingly, for larger perturbations $\delta$ the large-scale FSLE develops a minimum centred at approximately $\delta=1.1$ with $\lambda_{\rm LS}\approx 0.00428$ as shown in the inset, suggesting a large-scale predictability time scale of around $234$ time units. This is comparable to the decay time of autocorrelation of the slow variable $\tau_{\rm corr} \approx 208$, which is a measure for the transition times between the slow regimes. Since the large scale FSLE measures the predictability associated with transitions of the slow variable from one slow metastable state near $x^\star=\pm1$ to the other, this suggests that the trough in the FSLE spectrum is linked to the existence of slow regimes.\\ For values $\delta>1.5$ the perturbation size is comparable to the range of the slow variable which is approximately equal to $2.8$, rendering the FSLE meaningless.\\ 

\begin{figure}[]
\centering
\includegraphics[width = \columnwidth]{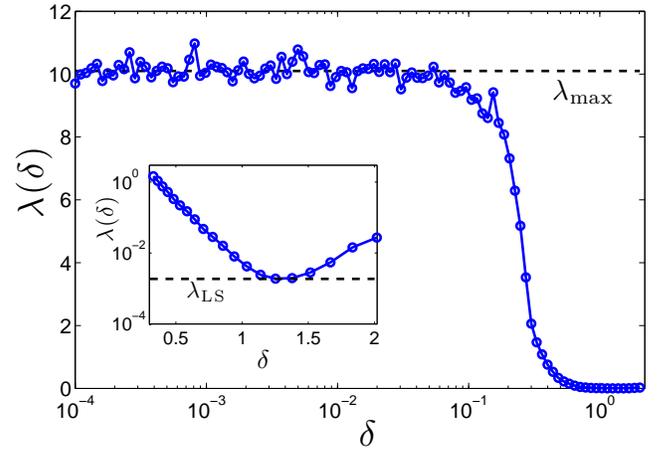}
\caption{FSLE $\lambda$ as a function of perturbation size $\delta$ for the model driven by the R\"ossler system (\ref{eqn:x_r})-(\ref{eqn:y3_r}) with slow regimes but without fast metastable regimes. The inset shows a zoom for large scale disturbances.}
\label{fig:FSLE_rossler}
\end{figure}

Figure \ref{fig:FSLE_rossler} shows the FSLE spectrum for the R\"ossler-driven system (\ref{eqn:x_r})-(\ref{eqn:y3_r}). As in Figure~\ref{fig:FSLE}, the spectrum exhibits a plateau at small scales corresponding to the maximal Lyapunov exponent $\lambda_{\rm max} \approx 10.1$. Most notably for the R\"ossler-driven system which does not support regimes, the large peaks at small perturbation amplitudes $\delta$ that we observed for the Lorenz-driven system are absent.  Since the small scale FSLEs describe fluctuations within each of the slow metastable states, this suggests that peaks in the FSLE spectrum are suggestive of fast regimes.\\ As shown in the inset, for larger perturbations $\delta$ the FSLE monotonically decreases to reach a lower plateau with the large scale FSLE $\lambda_{\rm LS}\approx 0.00227$ at the minimum at $\delta\approx 1.25$, suggesting a large-scale predictability time scale of around $440$ time units. Here  $1/\lambda_{\rm LS}$ is not close to $\tau_{\rm corr}\approx 234$. This is again due to the inevitable inaccurate estimation of the mean ${\bar {y}}_2$ which leads to an asymmetric probability density for $x$ as seen in Figure~\ref{fig:pdf}. Hence the trajectory resides longer in one well, thereby increasing the large scale predictability time; the autocorrelation time $\tau_{\rm corr}$, however, is measured for lag times much smaller than the mean residence time.\\ 

\begin{figure}[]
\centering
\includegraphics[width = \columnwidth]{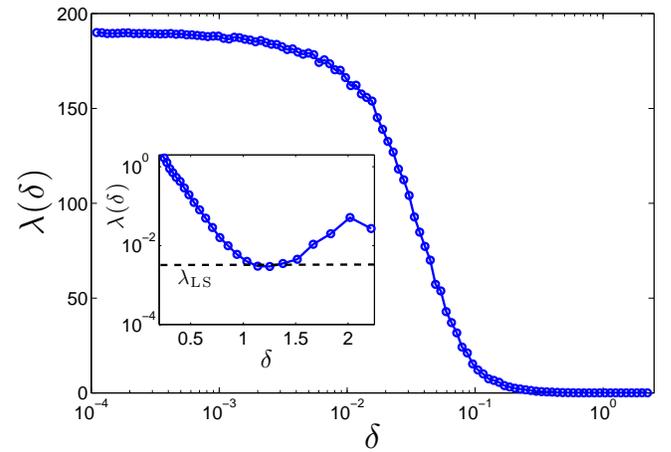}
\caption{FSLE $\lambda$ as a function of perturbation size $\delta$ for the climate model (\ref{eqn:climate}) with $\sigma^2=0.113$. The inset shows a zoom for large scale disturbances.
}
\label{fig:FSLE_climate}
\end{figure}

Figure \ref{fig:FSLE_climate} shows the FSLE spectrum for the stochastic model (\ref{eqn:climate}) with $\sigma^2 = 0.113$, which best approximates \cite{Mitchell11b} the full dynamics of the Lorenz-driven system (\ref{eqn:x})-(\ref{eqn:y3}). 
For small $\delta$, the FSLEs of the stochastic system do not reproduce the values for their parent systems, and we observe no small-scale plateau in the spectrum. This is expected as the maximal Lyapunov exponent $\lambda_{\rm max}$ is not defined for the stochastic system. However, at large scales $\delta$ (depicted in the inset) we find a minimum of the FSLE spectrum at $\delta\approx1.1$ with $\lambda_{\rm LS} = 0.00485$ for $\sigma^2 = 0.113$. This gives a large scale predictability time of $206$ time units, close to that obtained for the full parent model (\ref{eqn:x})-(\ref{eqn:y3}). As for the R\"ossler system there are no large peaks in the FSLE spectrum.\\

From these numerical simulations we now formulate our main hypothesis which we corroborate in the forthcoming Section by quantitative analytical theory and further simulations. We propose that the non-monotonicity observed in the FSLE spectra is due to the presence of regimes. In particular, large-scale troughs in the FSLE spectrum are an indication of slow regimes whereas small-scale peaks are caused by fast regimes. 


\section{Non-monotonicity of FSLE spectra for systems involving regimes: theory}
\label{sec:theory}

We now explain the numerical observations of the previous Section and relate them to the existence of slow and fast regimes, respectively. The minima at large scales will be explained by calculating most likely trajectory separations supported by a bimodal probability density function. The large peaks at small scales will be explained by a simple heuristic argument involving rapid switches of the fast dynamics between lobes of the Lorenz attractor. We denote the perturbation size corresponding to the minimum of the FSLE spectrum associated with slow regimes by $\delta_{\rm S}$. Similarly, we denote by $\delta_{\rm F}$ the perturbation size corresponding to the peaks associated with fast regimes.


\subsection{FSLE spectra for slow regimes}
\label{sec:slowregime}

The observed minimum of the FSLE spectrum $\lambda(\delta)$ at large scales can be understood by considering the bimodal probability density function of the slow variables. Before deriving an analytic expression for the large scale perturbation size $\delta_{\rm S}$, we give a heuristic argument why a minimum in the FSLE spectrum occurs for multimodal probability density functions. As seen in Figure~\ref{fig:pdf}, each of the two maxima in the probability density function has a characteristic width of roughly $1.25$. Perturbations larger than this size therefore likely correspond to a pair of trajectories with members residing in opposite wells of the potential $V(x)$. Perturbations smaller than this size likely correspond to a pair of trajectories with members residing in the same potential well. Hence there should exist a separation $\delta_{\rm S}$ with associated error growth rate $\lambda_{\rm LS}$ such that perturbations smaller than $\delta_{\rm S}$ will separate quicker, being pulled towards their mutual potential minimum, and perturbations slightly larger will separate quicker as they are pulled towards their respective closest potential minima.\\

We quantify this phenomenological argument by estimating the most likely configurations of pairs of trajectories which are separated by $\delta$. We denote the values of the slow variable $x$ of a pair of trajectories by $\xi$ and $\eta$. Let $p(\xi,\eta)$ be the joint probability function for two trajectories which were initially separated by $\delta_0$ to assume state values $\xi$ and $\eta$ respectively. The state values of the pair of trajectories $\xi$ and $\eta$ are then random variables drawn from this joint probability $p(\xi,\eta)$. For sufficiently large separations the two trajectories will have decorrelated and we can treat $\xi$ and $\eta$ as statistically independent, and approximate $p(\xi,\eta)=p(\xi)p(\eta)$. We have numerically verified this assumption for sufficiently large $\delta$. We perform the expectation value analytically utilizing the reduced stochastic model (\ref{eqn:climate}) and its invariant density (\ref{e.rhohat}) and set $p(x)=\hat\rho(x)$. This is justified by the theorems which underpin stochastic homogenization 
\cite{PavliotisStuart,MelbourneStuart11} as well as our numerical observations (cf. Figure~\ref{fig:pdf}) which state that the statistics of the full deterministic system converges to the statistics of the reduced stochastic system for $\varepsilon\to 0$.

The expectation value $\Xi$ of a location $\xi$ conditioned on all possible pairs which are separated by $\delta$ is given by
\begin{align}
\nonumber \Xi(\delta) 
&= \frac{1}{\cal{Z}}\int_0^\infty\int_0^\infty d\xi d\eta \;\xi\, p(\xi,\eta) {\bm \delta}(|\xi-\eta|-\delta)\\
&\approx \frac{1}{\cal{Z}}\int_0^\infty d\xi \; \xi \, \hat\rho(\xi)\left(\hat\rho(\xi+\delta)+\hat\rho(\xi-\delta)  \right)
\; ,
\label{eqn:Xi}
\end{align}
where $\cal Z$ is the normalization constant, and the bold-face $\bm \delta$ denotes the Dirac $\delta$-function. We only consider positive values of $\xi$, justified by the symmetry of our problem. As approximations we assumed statistical independence of $\xi$ and $\eta$, and ignored the conditioning of the expectation value on the initial separation $\delta_0$. 

In the case of a bimodal probability density function, $\Xi(\delta)$ will decrease initially with increasing separation $\delta$ allowing the pair of trajectories $(\xi,\eta)$ to arrange themselves within one of the two potential wells. For sufficiently large separations $\delta$, however, $\Xi$ will increase linearly with $\delta$ and the two trajectories will be in opposite wells. The curve of $\Xi(\delta)$ obtains its minimum when $\delta \approx b$.

The functional form of the asymptotic linear behaviour of the expectation value $\Xi(\delta)$ for large $b$ can be determined by expanding the probability density function $\hat{\rho}(x)$ around the maxima at $x^\star=\pm 1$ with
\[
\hat{\rho}(x) \sim \exp{\left(-\frac{2ab^2(x-b)^2}{\sigma^2} \right)} + \exp{\left(-\frac{2ab^2(x+b)^2}{\sigma^2} \right)}
\; .
\]
Upon inserting this approximation into (\ref{eqn:Xi}) a lengthy but straightforward calculation yields the asymptotic behaviour $\Xi(\delta) \rightarrow \delta / 2$ in the limit of large perturbation sizes $\delta \rightarrow \infty$.

Following our heuristic argument from above we may define the large-scale error perturbation $\delta_{\rm S}$ corresponding to maximal predictability as the value for which $\Xi(\delta)$ assumes its (unique) minimum. However, we find better numerical agreement if we estimate $\delta_{\rm S}$ as the value of $\delta$ for which $\Xi(\delta)$ is sufficiently close to its asymptotic behaviour, i.e. $\delta_{\rm S}$ solves
\begin{align}
\frac{ \Xi(\delta_{\rm S}) - \delta_{\rm S} / 2}{\Xi(\delta_{\rm S})} = \theta\; ,
\label{XideltaS}
\end{align}
where we chose $\theta=0.01$. The two definitions become indistinguishable for large values of $b$.

\begin{figure}[]
\centering
\includegraphics[width = 0.8\columnwidth]{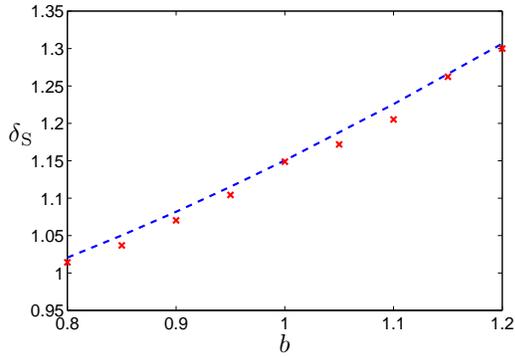}
\caption{Perturbation size $\delta_{\rm S}$ associated with the large-scale minimum of the FSLE spectrum as a function of the separation $b$ of the potential minima of $V(x)$ for fixed values of $a=1$ and $k=4/90$. The crosses denote values obtained by averaging 20000 simulations of the Lorenz-driven system (\ref{eqn:x})-(\ref{eqn:y3}); the dashed line is our analytical prediction (\ref{XideltaS}).}
\label{fig:deltaS1}
\end{figure}

\begin{figure}[]
\centering
\includegraphics[width = 0.8\columnwidth]{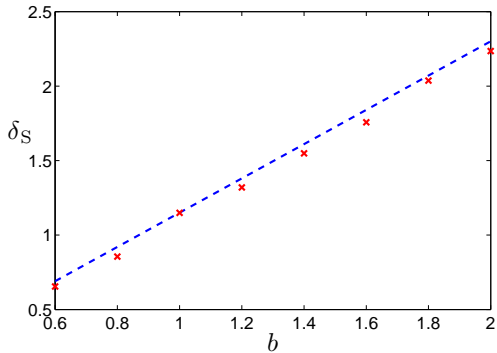}
\caption{As in Figure \ref{fig:deltaS1}, but with $k=(4/90)b^2$.}
\label{fig:deltaS2}
\end{figure}

To test our analytical prediction we now vary the parameters $a$ and $b$ of the potential $V(x)$ in (\ref{V}), which measure the height of the potential well and separation of the potential minima. In Figure~\ref{fig:deltaS1} we show a comparison between $\delta_{\rm S}$ as calculated from estimating the FSLE spectra using numerical simulations of the dynamics of the Lorenz-driven system (\ref{eqn:x})-(\ref{eqn:y3}) and our analytical prediction (\ref{XideltaS}), showing good agreement. Note that the behaviour is almost linear (however, $b$ not only affects the distance between the minima at $x^\star\pm1$ but also the potential well height). Linear scaling with $b$ can be achieved if we scale the coupling $k$ with $b$ according to $k\to b^2\sqrt{a}k$, such that the probability density function $\hat \rho$ is invariant upon scaling $x\to b x$ (cf. (\ref{e.rhohat}) with the definition of $\sigma^2$ in (\ref{e.sigma2})), as illustrated in Figure~\ref{fig:deltaS2}. We have tested that $\delta_{\rm S}$ is insensitive to i) varying $a$ for fixed $b$ and also to ii) varying the coupling $k$ for fixed $b$, consistent with our formula (\ref{XideltaS}) (not shown).\\

Hence, slow regimes and fast transitions between them cause the FSLE spectrum to exhibit a distinct minimum at large error perturbation sizes, with error growth rate related to the average residence time in each potential well (or decay rate of the autocorrelation time).

%






\subsection{FSLE spectra for fast regimes}
\label{sec:fastregime}

\begin{figure}[]
\centering
\includegraphics[width = \columnwidth]{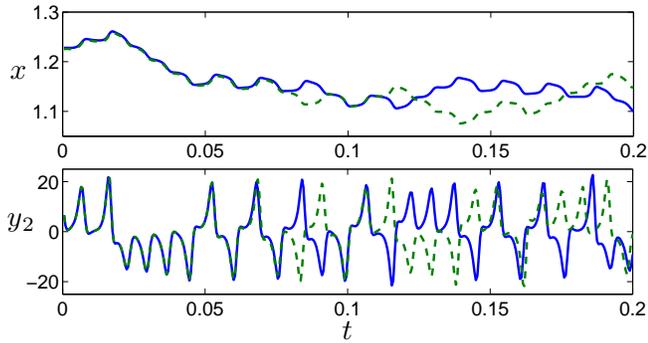}
\caption{Two trajectories obtained from integrating the Lorenz-driven system (\ref{eqn:x})-(\ref{eqn:y3}) from nearby initial conditions with separation $10^{-3}$, showing short time dynamics of the slow $x$ variable, and how increments correlate with switches between regimes of the $y_2$-component of the Lorenz-63 subsystem, corresponding to the lobes of the butterfly attractor.}
\label{fig:xy2_traj}
\end{figure}

We now present a simple heuristic argument explaining the observed peaks in the FSLE spectrum $\lambda(\delta)$ for the Lorenz-driven system (\ref{eqn:x})-(\ref{eqn:y3}). We link these to the presence of regimes in the fast process and the switching of the fast dynamics between the two lobes of the butterfly attractor. Figure \ref{fig:xy2_traj} depicts the slow $x$ and fast $y_2$ variables of two typical trajectories which are used to calculate the FSLE. The slow variable $x$ evolves in a step-like fashion with step size $\Delta x$. 
Separations between nearby trajectories therefore occur in ``units'' of $\Delta x$. Separations of $\delta = m \Delta x$ can only occur when the $y_2$ components of each trajectory are on opposite lobes of the Lorenz attractor.  We measured the period of one (fast) revolution around a lobe of the Lorenz attractor to be $T_{f}\approx 0.00691$. Integrating the slow dynamics (\ref{eqn:x}) over one fast period $T_{f}$ within a lobe and assuming that no transitions between slow metastable states at $x^* = \pm 1$ occur so that $\int_0^{T_{f}} x(t)(1 - x^2(t)) dt = 0$, we can approximate the step size $\Delta x$ by
\begin{align*}
\Delta x = \frac{k}{\eps} T_{f} \left<|y_2|\right>_t \; ,
\end{align*}
where $\left< |y_2| \right>_t = \frac{1}{T_{f}} \int_0^{T_{f}} |y_2| dt$. Hence we estimate
\begin{align}
\delta_{\rm F} = |\Delta x| \sim \frac{k}{\varepsilon}\; .
\label{deltaF}
\end{align}

We numerically obtain $\left< |y_2| \right>_t \approx 10$ as the average value of $|y_2|$ in each of the fast regimes, i.e. the lobes of the butterfly attractor. For our parameters with $\varepsilon^2=0.01$ and $k=4/90$, this yields a step size of $\Delta x \approx 0.0307$ which corresponds reasonably well with the observed location of the first peak in the FSLE spectrum in Figure~\ref{fig:FSLE} at $\delta_{\rm F}=0.0278$. The location of the second peak at $\delta_{\rm F}=0.0790$ is roughly approximated by $2\Delta x = 0.0614$ according to the above argument. 
The corresponding FSLEs $\lambda(\delta)$ can be estimated as follows. We assume that the separation of slow $x$ trajectories over short times $t ={\cal O}(T_{f})$ is approximately linear, and the initial separation $\delta$ of trajectories prior to taking a ``step'' in opposite directions is small. We define the times $t_m$ and $t_{rm}$ it takes for trajectories to separate by $m\Delta x$ and $rm \Delta x$ respectively. Assuming trajectories initially are infinitesimally separated and subsequently move apart at a constant rate of ${2\Delta x}/{T_{f}} = {m\Delta x}/{t_m} = {rm \Delta x}/{t_{rm}}$ (where $2\Delta x$ is the separation of trajectories after one step taken in opposite directions, see Figure \ref{fig:xy2_traj}), the time $T_r(m\Delta x)$ taken for a perturbation of size $m\Delta x$ to grow to size $rm\Delta x$ is 
\begin{align*}
T_r(m\Delta x) = t_{rm} - t_m = \frac{m(r-1)T_{f}}{2}\; ,
\end{align*}  
and so we can approximate the FSLE for the $m$th separation $m \Delta x$ using (\ref{e.FSLE}) as
\[
\lambda(m \Delta x) = \frac{2 \ln(r)}{m(r-1)T_{f}}\;.
\]
For $r = 1.1$ we find $\lambda(\Delta x) = 276$ and $\lambda(2\Delta x) = 138$, which, given the crude approximations, provide a reasonable estimate of the numerically observed peaks $\lambda(0.0278) = 394$ and $\lambda(0.0790) = 104$ in Figure \ref{fig:FSLE}.\\

\begin{figure}[]
\centering
\includegraphics[width = 0.8\columnwidth]{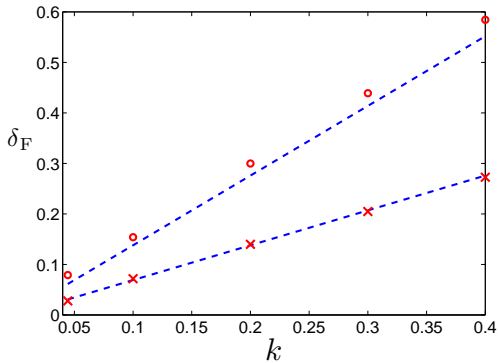}
\caption{Perturbation size $\delta_{\rm F}$ associated with peaks in the FSLE spectrum as a function of the coupling $k$ obtained from simulations of the Lorenz-driven system (\ref{eqn:x})-(\ref{eqn:y3}). The crosses denote the locations of the first peak, the circles denote the location of the second peak (cf. Figure~\ref{fig:FSLE}); the dashed lines are our analytical predictions (\ref{deltaF}).}
\label{fig:changek}
\end{figure}

According to our analytical expression for the location of the peaks (\ref{deltaF}), $\delta_{\rm F}$ scales linearly with the coupling parameter $k$. This is confirmed in Figure~\ref{fig:changek} where we show $\delta_{\rm F}$ as a function of $k$ obtained from numerical simulations of the Lorenz-driven system (\ref{eqn:x})-(\ref{eqn:y3}). We have checked (not shown) that $\delta_{\rm F}$ is insensitive to changes in $a$ and $b$ for fixed $k$ which would only affect the slow regimes.

%




\section{Discussion}
\label{sec:discussion}

We have studied the dependency of error growth rates on the amplitude of the initial error for a multiscale toy model with slow metastable states and fast regimes by calculating the finite size Lyapunov exponents. We found that the error growth rates can be a highly non-monotonic function of the initial error size in the presence of regimes. In particular we found that slow regimes produce minima in the FSLE spectrum at large scales, indicating enhanced predictability. On the other hand, fast regimes in the dynamics produce regions of rapid divergence of trajectories of the slow degrees of freedom, indicating poor predictability at those scales. This loss in predictability is found to be far greater than expected from the maximal Lyapunov exponent. In the context of ensemble generation, either for ensemble forecasts or for data assimilation, this means that there are initial perturbation sizes which may experience stronger amplification than infinitesimal perturbations along the most unstable eigendirections corresponding to the maximal Lyapunov exponent. Simple analytical arguments were employed to calculate the respective predictability times and critical perturbation sizes. Stochastic parametrizations of the fast processes do not exhibit peaks in the FSLE spectrum but as effective models of the slow dynamics share the large-scale minima of the FSLE spectrum. The sensitivity of error growth rate in the presence of regimes suggests caution is required when generating ensembles for forecasts or when assimilating data on systems with regimes. It is pertinent to mention that the signatures in the FSLE spectrum of slow and fast regimes occur only if perturbations are taken of the slow variables only; if one were to measure perturbations over all variables or of only the fast variables, the FSLE spectrum would be dominated by the strongly chaotic behaviour of the fast variables without any large-scale troughs or large peaks.\\ 

Non-monotonous behaviour of the FSLE has been previously reported \cite{Torcini95,Cencini02}. Discrete maps involving singular derivatives such as the circle map are simple examples where finite size perturbations can grow faster than infinitesimal perturbations. Here we discussed the occurrence of narrow well-defined peaks, caused by the fast dynamics switching regimes. We note that such behaviour was not seen by \citet{Boffetta98}, where a system of nonlinearly coupled fast and slow Lorenz-63 systems was studied. This is entirely due to the nature of the coupling used for which the fast dynamics does not induce rapid variations in the slow dynamics; we have checked that for linear skew coupling large peaks in the FSLE spectrum are again observed. Linearly coupled Lorenz-63 systems were for example used to model ENSO events \cite{Pena04}. However, the example of \citet{Boffetta98} shows that the existence of fast regimes is not sufficient for the occurrence of peaks in the FSLE spectrum.\\

Multimodal probability density functions are not necessary for the existence of metastable regimes \cite{Wirth01,Majda06}. 
\citet{Boffetta98} found a minimum in the FSLE spectrum for a coupled map system which has a unimodal probability density function but dynamics which consists of laminar phases interrupted by intermittent large amplitude bursts. This is in accordance with our reasoning of a critical perturbation size above which the dynamics dramatically increases sensitivity. Further work is required to determine how well our arguments transfer to unimodal probability density functions.


We remark that our numerical results were performed by estimating the FSLE spectra using the algorithm proposed by \citet{Aurell97} and \citet{Boffetta98}. Our analytical results, however, do not employ the particular method used to calculate the FSLEs, and we expect the observed non-monotonous behaviour of the FSLE due to slow and fast regimes to hold when other algorithms \cite{Boffetta02} are used to estimate the FSLEs. We further remark that for higher dimensional slow subspaces the choice of norm used to measure separations and to normalize bred vectors was shown to significantly alter their statistical properties \cite{PrimoEtAl2005,HallerbergEtAl2010}. Again, the generality of the arguments used here suggests that the choice of norm will not alter the occurrence of the non-monotonicity of the growth rates. This is planned for further research.


\begin{acknowledgments}
We would like to thank Armin K\"ohl for valuable discussions on the North Atlantic Oscillation, and Jeffrey Kepert for pointing us to the work of \citet{Nguyen11}. GAG acknowledges support from the Australian Research Council. LM acknowledges support from an Australian Postgraduate Award, and the University of Vermont Advanced Computing Centre.
\end{acknowledgments}



\begin{thebibliography}{}

\bibitem[Aurell {\em et~al.}(1997)Aurell, Boffetta, Crisanti, Paladin, and
  Vulpiani]{Aurell97}
Aurell, E., Boffetta, G., Crisanti, A., Paladin, G., and Vulpiani, A. (1997).
\newblock Predictability in the large: An extension of the concept of
  {L}yapunov exponent.
\newblock {\em Journal of Physics A: Mathematical and General\/}, {\bf 30}(1),
  1--26.

\bibitem[Boffetta {\em et~al.}(1998a)Boffetta, Giuliani, Paladin, and
  Vulpiani]{Boffetta98}
Boffetta, G., Giuliani, P., Paladin, G., and Vulpiani, A. (1998a).
\newblock An extension of the {L}yapunov analysis for the predictability
  problem.
\newblock {\em Journal of the Atmospheric Sciences\/}, {\bf 55}(23), 3409 --
  3416.

\bibitem[Boffetta {\em et~al.}(1998b)Boffetta, Crisanti, Paparella, Provenzale,
  and Vulpiani]{Boffetta98b}
Boffetta, G., Crisanti, A., Paparella, F., Provenzale, A., and Vulpiani, A.
  (1998b).
\newblock Slow and fast dynamics in coupled systems: A time series analysis
  view.
\newblock {\em Physica D\/}, {\bf 116}(3-4), 301--312.

\bibitem[Boffetta {\em et~al.}(2002)Boffetta, Cencini, Falcioni, and
  Vulpiani]{Boffetta02}
Boffetta, G., Cencini, M., Falcioni, M., and Vulpiani, A. (2002).
\newblock Predictability: a way to characterize complexity.
\newblock {\em Physics Reports\/}, {\bf 356}(6), 367 -- 474.

\bibitem[Branstator and Berner(2005)Branstator and Berner]{BranstatorBerner05}
Branstator, G. and Berner, J. (2005).
\newblock {Linear and nonlinear signatures in the planetary wave dynamics of an
  AGCM: Phase space tendencies}.
\newblock {\em Journal of the Atmospheric Sciences\/}, {\bf 62}(1), 1792--1811.

\bibitem[Cai {\em et~al.}(2003)Cai, Kalnay, and Toth]{CaiKalnayToth03}
Cai, M., Kalnay, E., and Toth, Z. (2003).
\newblock Bred vectors of the {Z}ebiak-{C}ane model and their potential
  application to enso prediction.
\newblock {\em Journal of Climate\/}, {\bf 16}, 40--56.

\bibitem[Cencini and Torcini(2002)Cencini and Torcini]{Cencini02}
Cencini, M. and Torcini, A. (2002).
\newblock Linear and nonlinear information flow in spatially extended systems.
\newblock {\em Physical Review E\/}, {\bf 63}(5), 056201.

\bibitem[Charney and De~Vore(1979)Charney and De~Vore]{CharneyDeVore79}
Charney, J.~G. and De~Vore, J.~G. (1979).
\newblock {Multiple Flow Equilibria in the Atmosphere and Blocking}.
\newblock {\em Journal of the Atmospheric Sciences\/}, {\bf 36}(7),
  1205Ð--1216.

\bibitem[Deremble {\em et~al.}(2009)Deremble, D'Andrea, and
  Ghil]{DerembleEtAl09}
Deremble, B., D'Andrea, F., and Ghil, M. (2009).
\newblock Fixed points, stable manifolds, weather regimes, and their
  predictability.
\newblock {\em Chaos\/}, {\bf 19}, 043109.

\bibitem[Dijkstra(2005)Dijkstra]{Dijkstra}
Dijkstra, H. (2005).
\newblock {\em {Nonlinear Physical Oceanography: A Dynamical Systems Approach
  to the Large Scale Ocean Circulation and El Ni\~no}\/}.
\newblock Springer, 2nd edition.

\bibitem[Ditlevsen(1999)Ditlevsen]{Ditlevsen99}
Ditlevsen, P.~D. (1999).
\newblock Observation of $\alpha$-stable noise induced millennial climate
  changes from an ice-core record.
\newblock {\em Geophysical Research Letters\/}, {\bf 26}(10), 1441--1444.

\bibitem[D`Ovidio {\em et~al.}(2004)D`Ovidio, Fern{\'a}ndez,
  Hern{\'a}ndez-Garc{\'i}a, and L{\'o}pez]{DOvidio04}
D`Ovidio, F., Fern{\'a}ndez, V., Hern{\'a}ndez-Garc{\'i}a, E., and L{\'o}pez,
  C. (2004).
\newblock {Mixing structures in the {M}editerranean {S}ea from finite-size
  {L}yapunov exponents}.
\newblock {\em Geophysical Research Letters\/}, {\bf 31}(17), 1--4.

\bibitem[D`Ovidio {\em et~al.}(2009)D`Ovidio, Isern-Fontanet, L{\'o}pez,
  Hern{\'a}ndez-Garc{\'i}a, and Garc\'ia-Ladona]{DOvidio09}
D`Ovidio, F., Isern-Fontanet, J., L{\'o}pez, C., Hern{\'a}ndez-Garc{\'i}a, E.,
  and Garc\'ia-Ladona, E. (2009).
\newblock {Comparison between {E}ulerian diagnostics and finite-size {L}yapunov
  exponents computed from altimetry in the {A}lgerian basin}.
\newblock {\em Deep-Sea Research I\/}, {\bf 56}, 15--31.

\bibitem[Frauen and Dommenget(2010)Frauen and Dommenget]{Frauen10}
Frauen, C. and Dommenget, D. (2010).
\newblock {El Ni\~no and La Ni\~na amplitude asymmetry caused by atmospheric
  feedbacks}.
\newblock {\em Geophysical Research Letters\/}, {\bf 37}(18), L18801.

\bibitem[Froyland and Padberg(2009)Froyland and Padberg]{FroylandPadberg09}
Froyland, G. and Padberg, K. (2009).
\newblock {Almost-invariant sets and invariant manifolds Ñ {C}onnecting
  probabilistic and geometric descriptions of coherent structures in flows}.
\newblock {\em Physica D\/}, {\bf 238}, 1507--1523.

\bibitem[Givon {\em et~al.}(2004)Givon, Kupferman, and Stuart]{Givonetal04}
Givon, D., Kupferman, R., and Stuart, A. (2004).
\newblock Extracting macroscopic dynamics: Model problems and algorithms.
\newblock {\em Nonlinearity\/}, {\bf 17}(6), R55--127.

\bibitem[Greatbatch(2000)Greatbatch]{Greatbatch00}
Greatbatch, R.~J. (2000).
\newblock {The North Atlantic Oscillation}.
\newblock {\em Stochastic Environmental Research and Risk Assessment\/}, {\bf
  14}(4), 213--242.

\bibitem[Hallerberg {\em et~al.}(2010)Hallerberg, Pa\'zo, L\'opez, and
  Rodrigu\'ez]{HallerbergEtAl2010}
Hallerberg, S., Pa\'zo, D., L\'opez, J.~M., and Rodrigu\'ez, M.~A. (2010).
\newblock Logarithmic bred vectors in spatiotemporal chaos: Structure and
  growth.
\newblock {\em Physical Review E\/}, {\bf 81}, 066204.

\bibitem[Hern{\'a}ndez-Carrasco {\em et~al.}(2011)Hern{\'a}ndez-Carrasco,
  L{\'o}pez, Hern{\'a}ndez-Garc{\'i}a, and Turiel]{HernandezCarrasco11}
Hern{\'a}ndez-Carrasco, I., L{\'o}pez, C., Hern{\'a}ndez-Garc{\'i}a, E., and
  Turiel, A. (2011).
\newblock {How reliable are finite-size Lyapunov exponents for the assessment
  of ocean dynamics?}
\newblock {\em Ocean Modelling\/}, {\bf 36}(3-4), 208--218.

\bibitem[Kalnay(2002)Kalnay]{Kalnay}
Kalnay, E. (2002).
\newblock {\em {Atmospheric Modeling, Data Assimilation and Predictability}\/}.
\newblock Cambridge University Press, Cambridge.

\bibitem[K{\'a}rolyi {\em et~al.}(2010)K{\'a}rolyi,
  Pattanty{\'u}s-{\'A}brah{\'a}m, Kr{\'a}mer, J{\'o}zsa, and
  T{\'e}l]{Karolyi10}
K{\'a}rolyi, G., Pattanty{\'u}s-{\'A}brah{\'a}m, M., Kr{\'a}mer, T., J{\'o}zsa,
  J., and T{\'e}l, T. (2010).
\newblock {Finite-size {L}yapunov exponents: A new tool for lake dynamics}.
\newblock {\em Proceedings of the ICE Engineering and Computational
  Mechanics\/}, {\bf 163}(4), 251--259.

\bibitem[Kondrashov {\em et~al.}(2004)Kondrashov, Ide, and Ghil]{Kondrashov04}
Kondrashov, D., Ide, K., and Ghil, M. (2004).
\newblock Weather regimes and preferred transition paths in a three-level
  quasigeostrophic model.
\newblock {\em Journal of the Atmospheric Sciences\/}, {\bf 61}, 568--587.

\bibitem[Kwasniok and Lohmann(2009)Kwasniok and Lohmann]{KwasniokLohmann09}
Kwasniok, F. and Lohmann, G. (2009).
\newblock Deriving dynamical models from paleoclimatic records: {A}pplication
  to glacial millennial-scale climate variability.
\newblock {\em Physical Review E\/}, {\bf 80}(6), 066104.

\bibitem[Legras and Ghil(1985)Legras and Ghil]{LegrasGhil85}
Legras, B. and Ghil, M. (1985).
\newblock Persistent anomalies, blocking and variations in atmospheric
  predictability.
\newblock {\em Journal of the Atmospheric Sciences\/}, {\bf 42}, 433--471.

\bibitem[Lorenz(1969)Lorenz]{Lorenz69}
Lorenz, E.~N. (1969).
\newblock The predictability of a flow which possesses many scales of motion.
\newblock {\em Tellus\/}, {\bf 21}(3), 289--307.

\bibitem[Luzzatto {\em et~al.}(2005)Luzzatto, Melbourne, and
  Paccaut]{Luzzatto05}
Luzzatto, S., Melbourne, I., and Paccaut, F. (2005).
\newblock The {Lorenz} attractor is mixing.
\newblock {\em Communications in Mathematical Physics\/}, {\bf 260}, 393--401.

\bibitem[Majda {\em et~al.}(2006)Majda, Franzke, Fischer, and
  Crommelin]{Majda06}
Majda, A.~J., Franzke, C., Fischer, A., and Crommelin, D.~T. (2006).
\newblock {Distinct metastable atmospheric regimes despite nearly {G}aussian
  statistics: {A} paradigm model}.
\newblock {\em Proceedings of the National Academy of Sciences\/}, {\bf
  103}(22), 8309--8314.

\bibitem[Melbourne and Stuart(2011)Melbourne and Stuart]{MelbourneStuart11}
Melbourne, I. and Stuart, A. (2011).
\newblock A note on diffusion limits of chaotic skew-product flows.
\newblock {\em Nonlinearity\/}, {\bf 24}, 1361--1367.

\bibitem[Mitchell and Gottwald(2012)Mitchell and Gottwald]{Mitchell11b}
Mitchell, L. and Gottwald, G.~A. (2012).
\newblock Data assimilation in slow-fast systems using homogenized climate
  models.
\newblock {\em Journal of the Atmospheric Sciences\/}, {\bf 69}(4), 1359--1377.

\bibitem[Nguyen {\em et~al.}(2011)Nguyen, Reeder, Davidson, Smith, and
  Montgomery]{Nguyen11}
Nguyen, M.~C., Reeder, M.~J., Davidson, N.~E., Smith, R.~K., and Montgomery,
  M.~T. (2011).
\newblock {Inner-core vacillation cycles during the intensification of
  Hurricane Katrina}.
\newblock {\em Quarterly Journal of the Royal Meteorological Society\/}, {\bf
  137}(657), 829--844.

\bibitem[Pavliotis and Stuart(2008)Pavliotis and Stuart]{PavliotisStuart}
Pavliotis, G.~A. and Stuart, A.~M. (2008).
\newblock {\em {Multiscale Methods: Averaging and Homogenization}\/}.
\newblock Springer, New York.

\bibitem[Pe\~na and Kalnay(2004)Pe\~na and Kalnay]{Pena04}
Pe\~na, M. and Kalnay, E. (2004).
\newblock Separating fast and slow modes in coupled chaotic systems.
\newblock {\em Nonlinear Processes in Geophysics\/}, {\bf 11}(3), 319--327.

\bibitem[Primo {\em et~al.}(2005)Primo, Rodrigu\'ez, L\'opez, and
  Szendro]{PrimoEtAl2005}
Primo, C., Rodrigu\'ez, M.~A., L\'opez, J.~M., and Szendro, I. (2005).
\newblock Predictability, bred vectors, and generation of ensembles in
  space-time chaotic systems.
\newblock {\em Physical Review E\/}, {\bf 72}, 015201(R).

\bibitem[Roulston and Neelin(2000)Roulston and Neelin]{RoulstonNeelin00}
Roulston, M.~S. and Neelin, D. (2000).
\newblock The response of an {ENSO} model to climate noise, weather noise and
  intraseasonal forcing.
\newblock {\em Geophysical Research Letters\/}, {\bf 27}(22), 3723--3726.

\bibitem[Tew~Kai {\em et~al.}(2009)Tew~Kai, Rossi, Sudre, Weimerskirch,
  L{\'o}pez, Hern{\'a}ndez-Garc{\'i}a, Marsac, and Garcon]{TewKai09}
Tew~Kai, E., Rossi, V., Sudre, J., Weimerskirch, H., L{\'o}pez, C.,
  Hern{\'a}ndez-Garc{\'i}a, E., Marsac, F., and Garcon, V. (2009).
\newblock {Top marine predators track {L}agrangian coherent structures}.
\newblock {\em Proceedings of the National Academy of Sciences\/}, {\bf
  106}(20), 8245--8250.

\bibitem[Torcini {\em et~al.}(1995)Torcini, Grassberger, and Politi]{Torcini95}
Torcini, A., Grassberger, P., and Politi, A. (1995).
\newblock Error propagation in extended chaotic systems.
\newblock {\em Journal of Physics A\/}, {\bf 28}(16), 4533--4541.

\bibitem[Toth and Kalnay(1993)Toth and Kalnay]{TothKalnay93}
Toth, Z. and Kalnay, E. (1993).
\newblock Ensemble forecasting at {NMC}: The generation of perturbations.
\newblock {\em Bulletin of the American Meteorological Society\/}, {\bf 74},
  2317--2330.

\bibitem[Toth and Kalnay(1997)Toth and Kalnay]{TothKalnay97}
Toth, Z. and Kalnay, E. (1997).
\newblock Ensemble forecasting at {NCEP} and the breeding method.
\newblock {\em Monthly Weather Review\/}, {\bf 125}(12), 3297--3319.

\bibitem[Tucker(1999)Tucker]{Tucker99}
Tucker, W. (1999).
\newblock {The Lorenz attractor exists}.
\newblock {\em Comptes Rendus de l'Acad\'emie des Sciences - Series I -
  Mathematics\/}, {\bf 328}(12), 1197 -- 1202.

\bibitem[Wirth(2001)Wirth]{Wirth01}
Wirth, V. (2001).
\newblock {Detection of hidden regimes in stochastic cyclostationary time
  series}.
\newblock {\em Physical Review E\/}, {\bf 64}(1), 016136.

\bibitem[Woollings {\em et~al.}(2008)Woollings, Hoskins, Blackburn, and
  Berrisford]{Woolings08}
Woollings, T., Hoskins, B., Blackburn, M., and Berrisford, P. (2008).
\newblock {A new {R}ossby wave-breaking interpretation of the {N}orth
  {A}tlantic {O}scillation}.
\newblock {\em Journal of the Atmospheric Sciences\/}, {\bf 65}(2), 609--626.

\bibitem[Zavala-Garay {\em et~al.}(2003)Zavala-Garay, Moore, Perez, and
  Kleeman]{ZavalaGaray03}
Zavala-Garay, J., Moore, A.~M., Perez, C.~L., and Kleeman, R. (2003).
\newblock {The response of a coupled model of {ENSO} to observed estimates of
  stochastic forcing}.
\newblock {\em Journal of Climate\/}, {\bf 16}(17), 2827--2842.

\end{thebibliography}


\end{document}